\documentclass[letterpaper,journal]{IEEEtran}
\usepackage{amsmath,amsfonts}
\usepackage{algorithmic}
\usepackage{algorithm}
\usepackage{array}
\usepackage[caption=false,font=footnotesize,labelfont=rm,textfont=rm,subrefformat=parens]{subfig}
\usepackage{textcomp}
\usepackage{stfloats}
\usepackage{graphicx}
\usepackage{epsfig,epsf,color,balance,cite}
\usepackage{epstopdf}
\usepackage{url}
\usepackage{verbatim}
\usepackage{graphicx}
\usepackage{cite}
\usepackage{amsmath}
\usepackage{amssymb}
\usepackage{bm}
\usepackage{color}
\allowdisplaybreaks[3]
\allowdisplaybreaks
\title{\LARGE{Rotatable Antenna-Enabled Satellite Communication: Joint Design of Boresight Alignment and Beam Tracking}}

\author{Tiantian Ma, Beixiong Zheng,~\IEEEmembership{Senior Member,~IEEE}, Changsheng You,~\IEEEmembership{Senior Member,~IEEE},\\
Ruiqi Liu,~\IEEEmembership{Senior Member,~IEEE},
and Robert Schober,~\IEEEmembership{Fellow,~IEEE}

\thanks{
    Tiantian Ma and Beixiong Zheng are with the School of Microelectronics, South China University of Technology, Guangzhou 511442, China (e-mail:mitiantianma@mail.scut.edu.cn; bxzheng@scut.edu.cn).
   
    Changsheng You is with the Department of Electronic and Electrical Engineering, Southern University of Science and Technology, Shenzhen 518055, China, and also with the Shenzhen Key Laboratory of Optoelectronics and Intelligent Sensing, Shenzhen 518055, China (e-mail: youcs@sustech.edu.cn)
    
    Ruiqi Liu is with the Wireless and Computing Research Institute, ZTE Corporation, Beijing 100029, China (email: richie.leo@zte.com.cn).

    Robert Schober is with the Institute for Digital Communications, Friedrich-Alexander-University Erlangen-N$\ddot{\mathrm{u}}$rnberg (FAU), 91054 Erlangen, Germany (e-mail: robert.schober@fau.de).
    
}
}	

\begin{document}
\maketitle%\large 
%\vspace{-5cm}
\begin{abstract}
Low Earth orbit (LEO) satellite links experience rapid angular variation due to high orbital velocities, which causes severe beam misalignment and array gain degradation under conventional fixed-antenna architectures. 
In this letter, we propose a rotatable antenna (RA)-enabled LEO communication framework, where RA arrays are deployed at both the satellite and the ground node (GN) to exploit antenna boresight reconfiguration as an additional spatial degree-of-freedom (DoF) for maintaining directional alignment under high mobility. 
By leveraging the rank-one line-of-sight (LoS) channel structure inherent to satellite links, we derive closed-form solutions for the joint design of the transmit/receive beamforming and antenna boresight directions, revealing that optimal performance can be achieved via decoupled alignment across antennas with low computational complexity.
To enable practical operation under dynamic conditions, we further develop a channel estimation and beam tracking protocol that exploits the predictable satellite orbit to continuously update boresight directions with low training overhead. 
Simulation results demonstrate that the proposed RA-enabled design significantly outperforms fixed and random boresight baselines in terms of achievable rate and robustness to angular variations, highlighting the effectiveness of rotational spatial reconfiguration in high-mobility satellite communications.
\end{abstract}

\vspace{-0.5cm}

\vspace{-0.5cm}
\section{Introduction}
\vspace{-0.2cm}
With the proliferation and commercialization of fifth-generation (5G) technologies, communication networks are evolving toward ubiquitous connectivity and full coverage, accompanied by increasing throughput and user demands~\cite{Liu2026ITU}.
However, this ambitious vision poses significant challenges to existing terrestrial network architectures, which struggle to support communications in non-terrestrial and remote outdoor environments (e.g., deserts) due to inherent limitations in coverage and capacity~\cite{Zhu2025Dynamic}.
To address this issue, low Earth orbit (LEO) satellite communication systems have emerged as a promising solution
%Low Earth orbit (LEO) satellite communication systems have therefore emerged as a promising solution
to offer wide coverage and cost-effective deployment~\cite{Liu2018Space}.
However, satellite communications remain challenging due to rapid angular variations caused by high orbital velocities and long propagation distances.
To compensate for the resulting severe path loss, directional antennas and phased arrays are employed to form narrow and high-gain beams that maintain reliable links during satellite motion~\cite{Phased2022Chen,Li2022Downlink}.
%Existing studies typically exploit the predictable satellite motion to perform angle estimation and tracking via filtering techniques~\cite{Lin2024Exploiting}.
%Alternatively, position-based beam selection and heuristic update mechanisms are applied to enable beam tracking, which, however, fail to fully exploit the continuous spatial degrees of freedom and optimal beamforming design~\cite{Lee2025Dynamic}.
%Existing studies exploit the predictable satellite motion for beam tracking via filtering or position-based beam selection~\cite{Lin2024Exploiting,Lee2025Dynamic}.
%However, these techniques are typically implemented with fixed-antenna architectures and thus offer only limited spatial adaptability in such dynamic environments, leading to noticeable performance degradation as angular misalignment accumulates.
Nevertheless, conventional beamforming based on fixed-antenna architectures offer only limited spatial adaptability in such dynamic environments, leading to noticeable performance degradation as angular misalignment accumulates.
%However, these approaches rely on discrete or heuristic beam adaptation and fail to fully exploit continuous spatial degrees of freedom. 
%More importantly, they remain constrained by fixed-antenna architectures, whose limited spatial adaptability leads to severe performance degradation under accumulated angular misalignment.

%Recently, rotatable antenna (RA) architecture has emerged to overcome the above mentioned limitation by exploiting additional rotational degrees of freedom (DoFs), thereby substantially enhancing spatial flexibility and directional gain~\cite{Zheng2026Rotatable,Wu2025Modeling,Zheng2025Rotatable,Xiong2026Intelligent,Xiong2025Efficient}.
Recently, the rotatable antenna (RA) architecture has demonstrated the ability of exploiting additional rotational degrees-of-freedom (DoFs) to enhance spatial flexibility, thereby overcoming the directional misalignment inherent to fixed-antenna architectures for dynamic channels~\cite{Zheng2026Rotatable,Zheng2025Rotatable,Xiong2026Intelligent,zheng2026rotatableTuts}. 
Specifically, by enabling each antenna to mechanically or electronically adjust its orientation/boresight, RA can dynamically rotate its effective radiation pattern toward the instantaneous direction of the satellite, achieving additional boresight alignment gain beyond conventional beamforming based on fixed-antenna architectures.
This capability is particularly attractive for satellite communications, where the line-of-sight (LoS) path dominates and the relative geometry evolves in a predictable manner over time.
Recent studies have explored the benefits of RA across various terrestrial applications, such as physical layer security (PLS)~\cite{Dai2025Rotatable},
%Dai2026Rotatable},
integrated sensing and communication (ISAC)~\cite{Zhou2025Rotatable},
%cell-free multiple-input-multiple-output (MIMO)~\cite{},
and cognitive radio~\cite{Tan2026Rotatable}.
However, RA-aided satellite communication has not been investigated in the literature yet, which further motivates this work.
%Despite these appealing advantanges, the potential of RA in LEO satellite communication systems remains largely unexplored in existing studies.

\begin{figure}[t]
	\centering
    \includegraphics[width=0.6\linewidth]{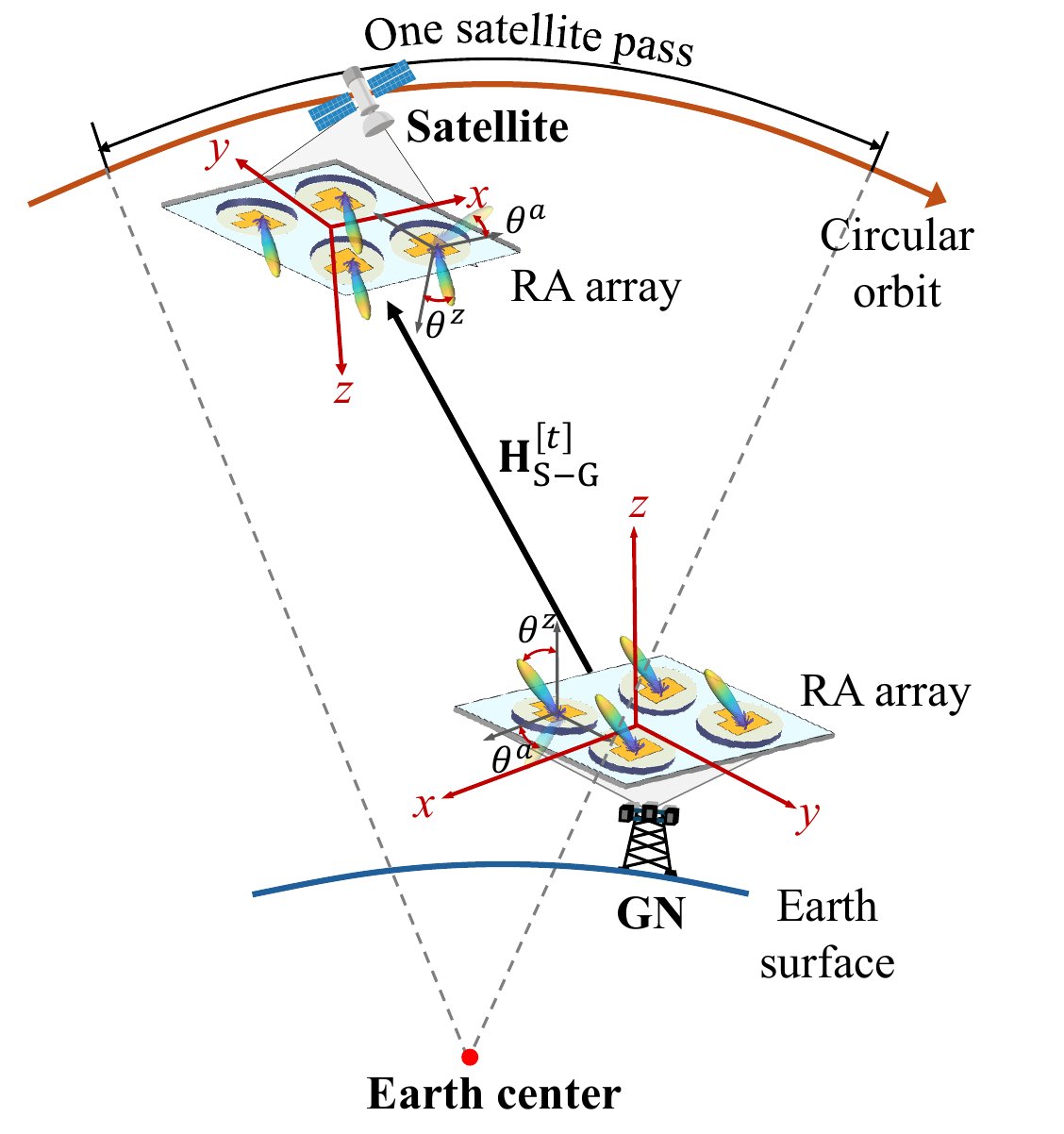}
    \vspace{-0.3cm}
	\caption{An illustration of LEO satellite communication with two-sided RA arrays.}
	\label{fig:system}
    \vspace{-0.6cm}
\end{figure}

Motivated by the above, we propose a two-sided RA-enabled LEO satellite communication system in this letter. By leveraging the rank-one far-field LoS channel structure, we jointly optimize the transmit/receive beamforming and the boresight directions of the RA arrays at both the ground node (GN) and the satellite to maximize the effective channel gain for data transmission.
Moreover, to handle the high mobility of the LEO satellite, the system employs a distributed channel estimation and beam tracking protocol that exploit the predictability of orbital trajectories to minimize training overhead. 
Numerical results demonstrate that the proposed scheme significantly outperforms various benchmark schemes.

%%By explicitly modeling the three-dimensional boresight direction of each RA and its directional gain pattern, we establish a far-field LoS channel model that captures the joint impact of antenna orientation, array response, and satellite motion. Building on this model, we show that the joint beamforming and boresight optimization problem can be decomposed into independent subproblems, each admitting a closed-form solution that aligns the antenna boresight with the instantaneous LoS direction under practical rotation constraints.
%Due to the high altitude of the satellite, the satellite to GN link can be well modeled as LoS in practice~\cite{You2020Massive}.
%Moreover, unlike other high mobility scenarios such as UAV communications, the trajectory of a LEO satellite is predictable, which can be exploited for efficient beam tracking and alignment~\cite{Zheng2022Intelligent}. Leveraging these characteristics, we jointly optimize the transmit and receive beamforming at the satellite and the GN, together with the orientation or boresight directions of each RA at both ends, to maximize the effective channel gain for downlink data transmission.
%Furthermore, we propose a distributed channel estimation and beam tracking framework tailored to RA-enabled LEO satellite systems to support the real-time boresight alignment under the high-mobility conditions while maintaining limited training overhead. Simulation results demonstrate that the proposed RA based scheme achieves a significantly higher achievable rate compared with conventional benchmarks.

\vspace{-0.5cm}
\section{System Model}
\vspace{-0.2cm}
\label{sec:model}
As shown in Fig.~\ref{fig:system}, we consider an RA-enabled LEO satellite communication system, where a GN communicates with a LEO satellite during the visible portion of its orbit (i.e., a satellite pass)\footnote{The single-GN case is considered as a baseline to highlight the fundamental performance gains of the proposed RA-enabled framework. The extension to multi-GN scenarios with inter-user interference management and multiuser beamforming is possible.}.
Particularly, both the GN and the satellite are equipped with uniform planar arrays (UPAs) comprising $N_{\mathrm G}$ and $N_{\mathrm S}$ RAs, respectively.
For notational simplicity, we let subscript $\mathrm X \in \{\mathrm G,\mathrm S\}$ indicate the considered node, where $\mathrm G$ and $\mathrm S$ represent the GN and the satellite, respectively.
Without loss of generality, we assume that the UPA at node X lies in the $x$-$y$ plane of a local Cartesian coordinate system (CCS) and is centered at the node reference position $\mathbf p_{\mathrm X}\in\mathbb{R}^{3\times 1}$, with $N_{\mathrm X}=N_{\mathrm X,x} \times N_{\mathrm X,y}$, where $N_{\mathrm X,x}$ and $N_{\mathrm X,y}$ denote the numbers of RAs along the $x$- and $y$-axes of the local CCS, respectively.
Moreover, the satellite is assumed to move along a circular orbit with radius $L_\mathrm{O}$ centered at the Earth’s center, while the GN is fixed on the Earth's surface.
\vspace{-0.5cm}
\subsection{Antenna Boresight Rotation}
%In this setting, the orientations of all RAs are modeled as adjustable parameters that can proactively reconfigure the effective channel.
Initially, the orientations/boresights of all RAs are assumed to be aligned with the $z$-axis of their corresponding local CCSs (see Fig.~\ref{fig:system}), and each RA is capable of independently adjusting its boresight direction in 3D space.
Such boresight adjustment can be described by a zenith angle $\theta^z$ and an azimuth angle $\theta^{a}$, as shown in Fig.~\ref{fig:system}.
The 3D boresight direction of the $i$-th RA at node X with $i\in\{1,\cdots, N_{\mathrm{X}}\}$, expressed in its local CCS, is then characterized by a unit pointing vector~\cite{Zheng2025Rotatable}
%To characterize the 3D orientation of each RA, the pointing vector associated with the zenith angle $\theta^z_{\mathrm X,i}$ and azimuth angle $\theta^a_{\mathrm X,i}$ of the $i$-th RA at node X is defined as
\begin{equation}
	\vec{\mathbf f}_{\mathrm X, i}=\left[\sin\theta^z_{\mathrm X,i}\cos\theta^a_{\mathrm X,i},\sin\theta^z_{\mathrm X,i}\sin\theta^a_{\mathrm X,i},\cos\theta^z_{\mathrm X,i}\right ]^{\mathrm T},
\end{equation}
where $(\cdot)^{\mathrm T}$ stands for the transpose operation.
%where $\mathbf R_{\mathrm X}$ denotes the rotation matrix associated with the RA array at node X.
%For the GN-side RA array, the array orientation is fixed and thus $\mathbf R_{\mathrm G}=\mathbf I$.
%In contrast, as the satellite moves along its orbit, the satellite-side RA array undergoes a corresponding rotation. Therefore, the rotation matrix of the RA array at the satellite is given by
%\begin{equation}
%    \mathbf R_{\mathrm S}(t)=\begin{bmatrix}
%        \cos{\phi_{\mathrm S}} & \cos{\theta_{\mathrm S}}\sin{\phi_{\mathrm S}} & -\sin{\theta_{\mathrm S}}\sin{\phi_{\mathrm S}}\\
%        0 & -\sin{\theta_{\mathrm S}} & -\cos{\theta_{\mathrm S}}\\
%        -\sin{\phi_{\mathrm S}} & \cos{\theta_{\mathrm S}}\cos{\phi_{\mathrm S}} & -\sin{\theta_{\mathrm S}}\cos{\phi_{\mathrm S}}
%    \end{bmatrix},
%\end{equation}
%where $\theta_{\mathrm S}=\theta_{\mathrm S}(t)$ and $\phi_{\mathrm S} = \phi_{\mathrm S}(t)$ denote the elevation and azimuth angle of the satellite at time $t$, respectively. 
For convenience, the pointing vectors of the RAs at the GN and the satellite are collected into matrices $\mathbf F_{\mathrm G}\triangleq[\vec{\mathbf f}_{\mathrm G, 1},\cdots, \vec{\mathbf f}_{\mathrm G, N_{\mathrm G}}] \in \mathbb{R}^{3\times N_{\mathrm G}}$ and $\mathbf F_{\mathrm S}\triangleq[\vec{\mathbf f}_{\mathrm S, 1},\cdots, \vec{\mathbf f}_{\mathrm S, N_{\mathrm S}}] \in \mathbb{R}^{3\times N_{\mathrm S}}$, respectively.

%Furthermore, to guarantee a sufficient boresight adjustment range and to mitigate mutual coupling between adjacent RAs, the zenith deviation of each RA from its reference normal is constrained by the maximum allowable zenith angle $\theta_{\mathrm{max}}$.
Furthermore, to account for practical rotational constraints, the zenith deviation of each RA from its reference boresight is constrained by the maximum allowable zenith angle $\theta_{\mathrm{max}}$.
Accordingly, feasible pointing vectors $\vec{\mathbf f}_{\mathrm X,i}$ must satisfy
\begin{equation}
\label{eq:range}
    0\le\arccos{(\vec{\mathbf f}_{\mathrm X,i}^{\mathrm T}{\mathbf e}_3)}\le\theta_{\mathrm{max}} \Leftrightarrow \vec{\mathbf f}_{\mathrm X,i}^{\mathrm T}{\mathbf e}_3\ge\cos{(\theta_{\mathrm{max}})},
\end{equation}
where $\mathbf e_3=[0,0,1]^{\mathrm T}$ represents the unit vector along the $z$-axis in the local CCS of each RA array.

\vspace{-0.5cm}
\subsection{Channel Model}
%Given the orientation adjustment capability, the effective antenna gain of each RA along a given propagation path is jointly determined by its orientation direction and the adopted antenna gain pattern.
In this letter, we adopt a generic directional gain pattern for each RA given by \cite{Balanis1996Antenna}
%Let $G(\epsilon,\varphi)$ denote the directional gain pattern adopted for each RA, which is expressed as follows~\cite{Balanis1996Antenna}.
\begin{equation}
\label{eq:pattern}
	G(\epsilon,\varphi) = 
	\begin{cases}
		G_{\mathrm{max}}\cos^{2p}(\epsilon), \quad
		&\epsilon \in [0,\frac{\pi}{2}),\varphi \in [0,2\pi) \\
		0, &\mathrm{otherwise,}
	\end{cases}
\end{equation}
where $(\epsilon,\varphi)$ denote incident angles of the signal with respect to the RA’s orientation direction $\vec{\mathbf f}$, 
%i.e., $\epsilon=\arccos{(\vec{\mathbf f}^{\mathrm T}\vec{\mathbf u})}$, 
$p$ denotes the directivity factor of the antenna, and $G_{\mathrm{max}}=2(2p+1)$ is the maximum boresight gain satisfying the power conservation law.
%Accordingly, the directional gain pattern coefficient vector of the RA array at node $\mathrm X$ is defined as
%\begin{equation}
%    \mathbf g_{\mathrm X}(\mathbf F_{\mathrm X})=[\sqrt{g(\vec{\mathbf f}_{\mathrm X, 1})},\cdots, \sqrt{g(\vec{\mathbf f}_{\mathrm X, N_{\mathrm X}})}]^{\mathrm T} \in \mathbb{R}^{N_{\mathrm X}\times 1},
%\end{equation}
%where $g(\vec{\mathbf f}_{\mathrm X, i})=\kappa_{\max}(\vec{\mathbf f}_{\mathrm X, i}^{\mathrm T}\vec{\mathbf u}_{\mathrm X})^{2p}$, $N_{\mathrm X}$ denotes the number of RAs at node $\mathrm X$.
%Here, $\vec{\mathbf u}_{\mathrm X}$ denotes the propagation direction of the link as observed from node $\mathrm X$. Specifically,
%\begin{equation}
%    \vec{\mathbf u}_{\mathrm G}=\frac{\mathbf q_{\mathrm S}-\mathbf p_{\mathrm G}}{\lVert \mathbf q_{\mathrm S}-\mathbf p_{\mathrm G} \rVert},\quad \vec{\mathbf u}_{\mathrm S}=\frac{\mathbf p_{\mathrm G}-\mathbf q_{\mathrm S}}{\lVert \mathbf p_{\mathrm G}-\mathbf q_{\mathrm S} \rVert}.
%\end{equation}
Accordingly, the directional gain of the $i$-th RA at node X$\in\{\mathrm{G,S}\}$ is given by $g(\vec{\mathbf f}_{\mathrm X, i})=G_{\max}\cos^{2p}{(\epsilon_{\mathrm X,i})}$, where $\cos(\epsilon_{\mathrm X,i})=\vec{\mathbf f}_{\mathrm X, i}^{\mathrm T}\vec{\mathbf u}_{\mathrm X}$ is the projection between the boresight direction $\vec{\mathbf f}_{\mathrm X, i}$ and the link propagation direction $\vec{\mathbf u}_{\mathrm X}$, as observed from node X.
%Specifically, $\vec{\mathbf u}_{\mathrm G}=\frac{\mathbf p_{\mathrm S}-\mathbf p_{\mathrm G}}{\lVert \mathbf p_{\mathrm S}-\mathbf p_{\mathrm G} \rVert}$, $ \vec{\mathbf u}_{\mathrm S}=\frac{\mathbf p_{\mathrm G}-\mathbf p_{\mathrm S}}{\lVert \mathbf p_{\mathrm G}-\mathbf p_{\mathrm S} \rVert}$.
Based on this, the directional gain vector of the RA array at node X is defined as
\begin{equation}
\label{eq:array-gain}
    \mathbf g_{\mathrm X}(\mathbf F_{\mathrm X})=\left[\sqrt{g(\vec{\mathbf f}_{\mathrm X, 1})},\cdots, \sqrt{g(\vec{\mathbf f}_{\mathrm X, N_{\mathrm X}})}\right]^{\mathrm T} \in \mathbb{R}^{N_{\mathrm X}\times 1}.
\end{equation}

%Accordingly, the directional gain pattern coefficient vector of the RA array at node X is defined as
%\begin{equation}
%    \mathbf g_{\mathrm X}(\mathbf F_{\mathrm X})=[\sqrt{g(\vec{\mathbf f}_{\mathrm X, 1})},\cdots, \sqrt{g(\vec{\mathbf f}_{\mathrm X, N_{\mathrm X}})}]^{\mathrm T} \in \mathbb{R}^{N_{\mathrm X}\times 1},
%\end{equation}
%where $g(\vec{\mathbf f}_{\mathrm X, i})=G_{\max}(\vec{\mathbf f}_{\mathrm X, i}^{\mathrm T}\vec{\mathbf u}_{\mathrm X})^{2p}$, $\vec{\mathbf u}_{\mathrm X}$ denotes the propagation direction of the link as observed from node $\mathrm X$. Specifically,
%\begin{equation}
%    \vec{\mathbf u}_{\mathrm G}=\frac{\mathbf q_{\mathrm S}-\mathbf p_{\mathrm G}}{\lVert \mathbf q_{\mathrm S}-\mathbf p_{\mathrm G} \rVert},\quad \vec{\mathbf u}_{\mathrm S}=\frac{\mathbf p_{\mathrm G}-\mathbf q_{\mathrm S}}{\lVert \mathbf p_{\mathrm G}-\mathbf q_{\mathrm S} \rVert}.
%\end{equation}

%%Likewise, the directional gain pattern coefficient vector of the receive RA array at the satellite is given by
%%\begin{equation}
%%    \mathbf g_{\mathrm S}(\mathbf F_{\mathrm S})=[\sqrt{g_1(\vec{\mathbf f}_{\mathrm S, 1})},\cdots, \sqrt{g_1(\vec{\mathbf f}_{\mathrm S, M})}]^{\mathrm T} \in \mathbb{R}^{M\times 1},
%%\end{equation}
%%where $g_1(\vec{\mathbf f}_{\mathrm S, m})=\kappa_{\max}({\vec{\mathbf f}_{\mathrm S, m}}^{\mathrm T}\vec{\mathbf p}_{\mathrm G})^{2p}$, with $\vec{\mathbf p}_{\mathrm G}=\frac{\mathbf p_{\mathrm G}-\mathbf q_{\mathrm S}}{\lVert \mathbf p_{\mathrm G}- \mathbf q_{\mathrm S} \rVert}$ being the unit direction vector from the satellite to GN.

For the link from the satellite to the GN, we adopt a time-varying far-field LoS channel model to account for the high-speed relative motion and long propagation distance between the GN and the satellite, as discussed in~\cite{Zheng2022Intelligent}.
%Let $\mathbf H^{[t]}_{\mathrm{S-G}} \in \mathbb R^{N\times M}$ denote the baseband equivalent time-varying channel from the satellite to GN at time $t$.
%Under this far-field LoS model, $\mathbf H^{[t]}_{\mathrm{S-G}}$ is of rank-one.
For notational convenience, we first define a one-dimensional (1D) steering vector function for a generic uniform linear array (ULA) as 
\begin{equation}
    \mathbf e(\phi,N)\triangleq\left[1,e^{-j\frac{2\pi}{\lambda}\phi},\cdots,e^{-j\frac{2\pi}{\lambda}(N-1)\phi}\right]^{\mathrm T}\in \mathbb{C}^{N\times 1},
\end{equation}
where $j=\sqrt{-1}$ is the imaginary unit, $\lambda$ denotes the signal wavelength, $\phi$ denotes the constant phase-shift difference between the signals at two adjacent antennas, and $N$ denotes the number of antennas in the ULA.
%Let $\mathbf a_{\mathrm G}(\vartheta^{[t]}_{\mathrm G}, \varphi^{[t]}_{\mathrm G})$ and $\mathbf a_{\mathrm S}(\vartheta^{[t]}_{\mathrm S}, \varphi^{[t]}_{\mathrm S})$ denote the array response vectors associated with the angle-of-arrival/departure (AoA/AoD) pairs $(\vartheta^{[t]}_{\mathrm G}, \varphi^{[t]}_{\mathrm G})$ and $(\vartheta^{[t]}_{\mathrm S}, \varphi^{[t]}_{\mathrm S})$ of the GN and the satellite at time $t$, respectively.
%Under the far-field plane wave assumption, the array response vectors can be represented as the Kronecker product of two steering vectors corresponding to the horizontal and vertical dimensions.
Let $\mathbf a_{\mathrm X}(\vartheta^{[t]}_{\mathrm X}, \varphi^{[t]}_{\mathrm X})$ denote the array response vector associated with the angle-of-arrival/angle-of-departure (AoA/AoD) pair of node X at time $t$.
Under the far-field plane wave assumption, 
%the array response vectors at both nodes can be represented as the Kronecker product of two steering vectors corresponding to the horizontal and vertical dimensions, which 
the array response vector at node X is given by
\begin{align}
    \label{eq:arrayRes1}
    \mathbf a_{\mathrm X}(\vartheta^{[t]}_{\mathrm X}, \varphi^{[t]}_{\mathrm X}) =
    \mathbf e(\Delta_{\mathrm X}\cos\varphi^{[t]}_{\mathrm X}\cos\vartheta^{[t]}_{\mathrm X},N_{\mathrm X,x}) \nonumber \\
    \otimes \mathbf e(\Delta_{\mathrm X}\cos\varphi^{[t]}_{\mathrm X}\sin\vartheta^{[t]}_{\mathrm X},N_{\mathrm X,y}),
\end{align}
where $\otimes$ denotes the Kronecker product and $\Delta_{\mathrm X}$ denotes the antenna spacing of the RA array at node X.

Accounting for the array directional gain pattern in \eqref{eq:array-gain}, the effective array response vector of the RA array at node X is defined as
\begin{equation}
    \mathbf b_{\mathrm X}(\mathbf F^{[t]}_{\mathrm X};\vartheta^{[t]}_{\mathrm X}, \varphi^{[t]}_{\mathrm X}) = \mathbf g_{\mathrm X}(\mathbf F^{[t]}_{\mathrm X}) \odot \mathbf a_{\mathrm X}(\vartheta^{[t]}_{\mathrm X}, \varphi^{[t]}_{\mathrm X}),
\end{equation}
where $\odot$ denotes the Hadamard product, $\mathbf F_{\mathrm X}^{[t]}$ denotes the RA pointing matrix at node X at time $t$.
Accordingly, the instantaneous far-field LoS channel between the satellite and the GN at time $t$ is given by
\begin{align}
	\label{eq:channel}
    \mathbf H^{[t]}_{\mathrm{S-G}}=\rho_{\mathrm{S-G}}^{[t]}
    \mathbf b_{\mathrm S}(\mathbf F^{[t]}_{\mathrm S};\vartheta^{[t]}_{\mathrm S}, \varphi^{[t]}_{\mathrm S})
    \mathbf b_{\mathrm G}^{\mathrm T}(\mathbf F^{[t]}_{\mathrm G};\vartheta^{[t]}_{\mathrm G}, \varphi^{[t]}_{\mathrm G}),
\end{align} 
where $\rho_{\mathrm{S-G}}^{[t]}=\frac{\beta}{d^{[t]}_{\mathrm{S-G}}}e^{\frac{-j2\pi}{\lambda}d^{[t]}_{\mathrm{S-G}}}$ denotes the complex-valued channel path gain between the satellite and the GN, $d_{\mathrm{S-G}}^{[t]}$ represents the instantaneous propagation distance, and $\beta$ is the reference path gain at a distance of 1 meter (m).

\vspace{-0.3cm}
\section{Joint Beamforming and Boresight Design}

In this section, we aim to maximize the effective GN-satellite channel gain by jointly optimizing the beamforming and RA boresight directions of both nodes.
%under perfect real-time channel state information (CSI).
%%For brevity, we drop the time index $[t]$ in this section without causing any confusion.
%We adopt a block fading model, where the channel remains constant within each transmission block but may vary across different blocks due to satellite motion. Therefore, the time index $[t]$ can be omitted in this section without ambiguity.
We adopt a block-wise quasi-static channel model, where the channel remains approximately constant within each sufficiently short transmission block. This is justified by the fact that the variation of the propagation angles induced by satellite motion is negligible over a sufficiently short time interval.
Therefore, the time index $[t]$ is omitted in this section without ambiguity.

Let ${\mathbf w}_{\mathrm G} \in \mathbb{C}^{N_{\mathrm G} \times 1}$ and $\mathbf w_{\mathrm S} \in \mathbb{C}^{N_{\mathrm S} \times 1}$ denote the normalized transmit/receive beamforming vectors applied at the GN and the satellite, respectively.
%The transmit beamforming vector is expressed as $\tilde{\mathbf w}_{\mathrm G}=\sqrt{P}{\mathbf w}_{\mathrm G}$, with a constant transmit power $\lVert\tilde{\mathbf w}_{\mathrm G}\rVert^2=P$.
Combing them with the channel model in \eqref{eq:channel} yields the effective GN-satellite channel gain as
\begin{equation}
	\label{eq:gain}
	\gamma(\mathbf F_{\mathrm G},\mathbf w_{\mathrm G},\mathbf F_{\mathrm S},\mathbf w_{\mathrm S}) =\lvert\mathbf w_{\mathrm G}^{\mathrm H} \mathbf H_{\mathrm{S-G}} \mathbf w_{\mathrm S}\rvert^2,
\end{equation}
where we have $\lVert \mathbf w_{\mathrm G} \rVert ^2 = 1$ and $\lVert \mathbf w_{\mathrm S} \rVert ^2 = 1$ for normalization, and $(\cdot)^{\mathrm H}$ denotes the Hermitian transpose.
Accordingly, the optimization problem for system design is formulated as
\begin{subequations}
	\begin{align}
		\hspace{-0.35cm}\mathrm{(P1):}  &\max_{\mathbf F_{\mathrm G},\mathbf F_{\mathrm S},\mathbf w_{\mathrm G},\mathbf w_{\mathrm S}}\hspace{-0.2cm} &&\gamma(\mathbf F_{\mathrm G},\mathbf w_{\mathrm G},\mathbf F_{\mathrm S},\mathbf w_{\mathrm S}) \\
		&\hspace{0.65cm}\mathrm{s.t.} &&\lVert\mathbf w_{\mathrm X}\rVert^2=1, \forall \mathrm X\in\{\mathrm{G,S}\}, \\
        %&&\lVert\mathbf w_{\mathrm G}\rVert^2\le P, \\
       % & &&\lVert\mathbf w_{\mathrm S}\rVert^2\le 1, \\
        & && \lVert \vec{\mathbf f}_{\mathrm X,i} \rVert = 1, \forall i,\mathrm X\in\{\mathrm{G,S}\}, \label{cons:fn} \\
		& && \vec{\mathbf f}_{\mathrm X,i}^{\mathrm T}{\mathbf e}_3\ge\cos(\theta_{\max}), \forall i,\mathrm X\in\{\mathrm{G,S}\}.
		\label{cons:range}
	\end{align}
\end{subequations}
%It can be verified that problem (P1) is a non-convex optimization problem due to the unit norm constraints in \eqref{cons:fn} and the non-convex constraints \eqref{cons:range}. 

By exploiting the rank-one structure of the far-field LoS channel in \eqref{eq:channel}, the effective channel gain in (P1) can be naturally decomposed into two independent effective gains for the GN and the satellite, i.e.,
\begin{equation}
	\label{eq:decomposition}
	\gamma(\mathbf F_{\mathrm G},\mathbf w_{\mathrm G},\mathbf F_{\mathrm S},\mathbf w_{\mathrm S}) =
	\lvert\rho_{\mathrm{S-G}}
	\rvert^2
	\underbrace{\lvert \mathbf w_{\mathrm G}^{\mathrm H}\mathbf b_{\mathrm G}\rvert^2}_{\gamma_{\mathrm G}(\mathbf F_{\mathrm G}, \mathbf w_{\mathrm G})} \cdot
	\underbrace{\lvert \mathbf b_{\mathrm S}^{\mathrm H}\mathbf w_{\mathrm S}\rvert^2}_{\gamma_{\mathrm S}(\mathbf F_{\mathrm S}, \mathbf w_{\mathrm S})}.
\end{equation}
%where $\gamma_{\mathrm G}(\mathbf F_{\mathrm G}, \mathbf w_{\mathrm G})$ and $\gamma_{\mathrm S}(\mathbf F_{\mathrm S}, \mathbf w_{\mathrm S})$ denote the effective gains at the GN and the satellite sides, respectively.
%In this regard, (P1) can be effectively reduced to independently maximizing the effective GN-side gain $\gamma_{\mathrm G}(\mathbf F_{\mathrm G}, \mathbf w_{\mathrm G})$ and satellite-side gain $\gamma_{\mathrm S}(\mathbf F_{\mathrm S}, \mathbf w_{\mathrm S})$ without the loss of optimality.
This decomposition reveals that the global optimum of (P1) is achieved when each side independently maximizes its effective gain.
For any given pointing matrices $\mathbf F_{\mathrm G}$ and $\mathbf F_{\mathrm S}$, the optimal transmit/receive beamformers follow the maximum-ratio transmission/combination (MRT/MRC) solution, i.e.,
\begin{align}
	\label{eq:beamform}
	\mathbf w^{\star}_{\mathrm X}=\frac{\mathbf b^*_{\mathrm X}}{\lVert \mathbf b_{\mathrm X} \rVert}, \quad \forall \mathrm X\in\{\mathrm{G,S}\},
\end{align}
where $(\cdot)^*$ denotes the complex conjugate.

%For any given pointing matrices $\mathbf F_{\mathrm G}$ and $\mathbf F_{\mathrm S}$, it is known that the maximum-ratio transmission/combination (MRT/MRC) beamformer is the optimal transmit/receive beamforming solution for maximizing  $\gamma_{\mathrm G}(\mathbf F_{\mathrm G}, \mathbf w_{\mathrm G})$ and $\gamma_{\mathrm S}(\mathbf F_{\mathrm S}, \mathbf w_{\mathrm S})$, i.e.,
%\begin{align}
%	\label{eq:beamform}
%	\mathbf w^{\star}_{\mathrm G}=\frac{\mathbf b^*_{\mathrm G}}{\lVert \mathbf b_{\mathrm G} \rVert}, \quad
%	\mathbf w^{\star}_{\mathrm S}=\frac{\mathbf b^*_{\mathrm S}}{\lVert \mathbf b_{\mathrm S} \rVert}.
%\end{align}
Substituting the optimal beamformers $\mathbf w^{\star}_{\mathrm X}$ into $\gamma_{\mathrm X}(\mathbf F_{\mathrm X}, \mathbf w_{\mathrm X})$ yields the following effective channel gain expression
\begin{align}
    \gamma &=\lvert \rho_{\mathrm{S-G}} \rvert^2 \lVert \mathbf b_{\mathrm G} \rVert^2
    \lVert \mathbf b_{\mathrm S}\rVert^2, \nonumber \\
    &= \frac{\beta G_{\mathrm{max}}^2}{d_{\mathrm{S-G}}^2} 
    \sum_{i=1}^{N_{\mathrm G}} \lvert ({\vec{\mathbf f}_{\mathrm G, i}}^{\mathrm T}\vec{\mathbf u}_{\mathrm G})^{2p} \rvert \cdot
    \sum_{i=1}^{N_{\mathrm S}} \lvert ({\vec{\mathbf f}_{\mathrm S, i}}^{\mathrm T}\vec{\mathbf u}_{\mathrm S})^{2p} \rvert.      
\end{align}
Therefore, the original problem (P1) can be equivalently decomposed into $N_{\mathrm G}+N_{\mathrm S}$ sub-problems, each of which independently optimizes the pointing vector of one RA.

For RA $i$ at node X, the subproblem is given by
\begin{subequations}
	\begin{align}
		\mathrm{(P2):}  &\max_{\vec{\mathbf f}_{\mathrm X,i}} \quad (\vec{\mathbf f}_{\mathrm X,i}^{\mathrm T}\vec{\mathbf u}_{\mathrm X})^{2p} \\
		&\hspace{0.2cm}\mathrm{s.t.} \quad \hspace{0.2cm} \eqref{cons:fn},\eqref{cons:range}. \nonumber
	\end{align}
\end{subequations}
Since $p>0$, the optimal solution to problem (P3) is given as
\begin{subequations}
	\label{eq:optBoresight}
	\begin{align}
		&\mathbf \theta_{\mathrm X,i}^{a\star} = \mathrm{arctan2}({\vec{\mathbf u}_{\mathrm X}^{\mathrm T}\mathbf e_2},{\vec{\mathbf u}_{\mathrm X}^{\mathrm T}\mathbf e_1}),  \\
		&\mathbf \theta_{\mathrm X,i}^{z\star} = \min\{\arccos(\vec{\mathbf u}_{\mathrm X}^{\mathrm T}\mathbf e_3), \theta_{\max}\},
	\end{align}
\end{subequations}
where $\mathbf e_1 = [1,0,0]^{\mathrm T}$ and $\mathbf e_2 = [0,1,0]^{\mathrm T}$ denote the unit vectors along the $x$- and $y$-axes in the local CCS of each RA array, respectively.

\textit{Remark:} The above result reveals that under dominant LoS propagation, the joint beamforming and RA boresight design admits a decoupled solution where each RA aligns its boresight with the propagation direction. Consequently, full array gain is preserved through physical boresight alignment without increasing the number of radio frequency (RF) chains. This structure also reduces the original joint optimization to simple boresight updates, enabling efficient real-time operation and facilitating predictive beam tracking, as discussed next.
%This decoupled approach provides a low-complexity foundation for real-time operation. By reducing the joint design to boresight alignments, the system preserves maximum array gain with minimal computational overhead, which is critical for the predictive beam tracking discussed in the following section.
%%Accordingly, in order to jointly optimize the RAs' boresight directions and beamforming at each side for enhancing the data transmission, we only need to acquire the real-time AoA/AoD pairs of the GN and the satellite.

\vspace{-0.3cm}
\section{Channel Estimation and Beam Tracking}
\label{sec:estimation}
\begin{figure}[t]
	\centering
	\includegraphics[width=\linewidth]{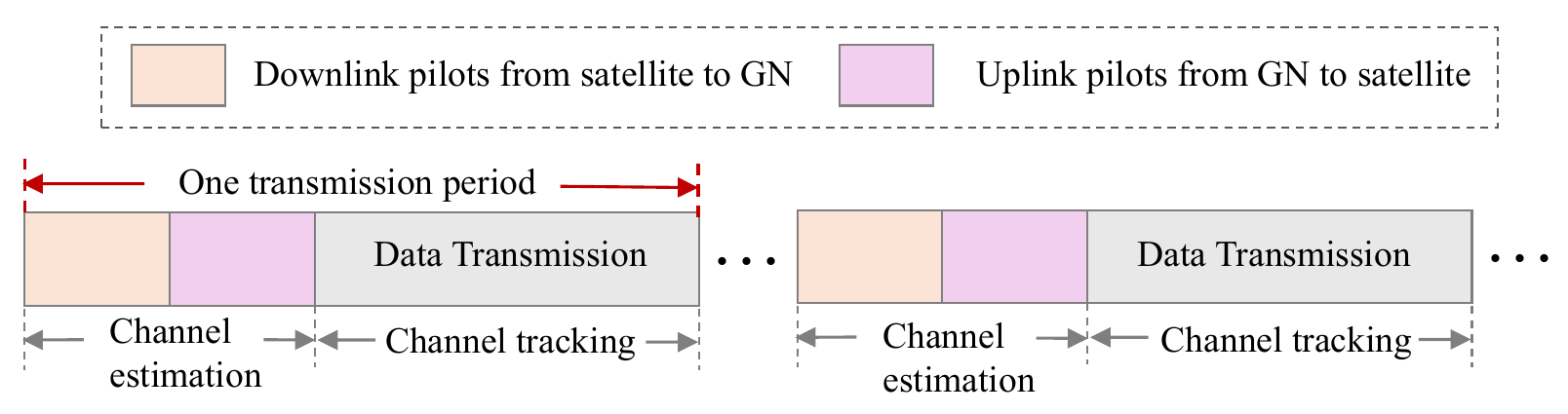}
	\caption{Channel estimation and tracking protocol for the proposed RA-enabled LEO satellite communication system.}
    \vspace{-0.6cm}
	\label{fig:protocal}
\end{figure}

As discussed in the previous section, the optimal boresight directions and beamforming vectors only depend on the local AoAs at each side rather than on the full channel matrix.
Motivated by this, we develop a distributed channel estimation and beam tracking scheme to obtain and continuously update the AoA parameters with low training overhead.
By leveraging these angular estimates, both GN and satellite can dynamically adjust their beamforming vectors and RA boresight directions to maintain directional alignment during data transmission.
The overall protocol, illustrated in Fig.~\ref{fig:protocal}, consists of two main procedures executed in an alternating manner.
In the following, we elaborate on the two procedures in detail.

%\begin{itemize}
%    \item \textbf{Channel estimation:} The satellite first transmits downlink pilots, from which the GN estimates its AoA pair $(\vartheta_{\mathrm G}, \varphi_{\mathrm G})$. Based on this estimate, the GN updates its beamforming vector and RA boresight directions according to \eqref{eq:beamform} and \eqref{eq:optBoresight}. The GN then sends the uplink pilots toward the satellite, enabling the satellite to estimate its own local AoA pair $(\vartheta_{\mathrm S}, \varphi_{\mathrm S})$ and update its beamforming vector and boresight directions in the same manner.
%    \item \textbf{Beam tracking:} After each channel training period, both the GN and the satellite predict the temporal evolution of their local AoA/AoD parameters by exploiting the known satellite orbital information.
 %   With these updated angles,the beamforming vectors and RA boresight directions are continuously updated during the transmission frame until the next channel estimation stage.
%\end{itemize}
%In the following, we elaborate on the two procedures in detail.

\vspace{-0.5cm}
\subsection{Channel Estimation}
In this subsection, we focus on the initial channel estimation procedure within each transmission period. 
Each channel training period consists of two stages, namely the downlink and uplink training blocks, containing $L_{\mathrm D}$ and $L_{\mathrm U}$ pilot symbols, respectively.
We consider a practically short duration for each training stage, during which the satellite-GN channel is assumed to remain approximately constant.
%For notational simplicity, the time index $[t]$ is omitted in this subsection.

During the downlink training stage, the satellite transmits pilot symbols to enable the GN to estimate the local AoA pair.
%%where we fix the boresight directions of RAs and the beamforming vector at the satellite side, and dynamically adjust the boresight directions of RAs at the GN over different pilot symbols to facilitate the channel estimation at the GN side.
%The satellite employs a fixed transmit beamforming vector and RA boresight directions during this stage, while the GN adjusts the boresight directions of its RAs across pilot symbols to facilitate angle estimation.
Both the satellite and the GN employ fixed transmit beamforming vectors and RA pointing matrices during the training stage, which are inherited from the previous transmission period.
Based on the channel model in \eqref{eq:channel}, the received signal at the GN is expressed as
\begin{equation}
	\mathbf y_{\mathrm G}^{(l)} = \sqrt{P}\mathbf H_{\mathrm{S-G}}\mathbf w_{\mathrm S}s^{(l)}+\mathbf n^{(l)}, \quad l=1,\cdots,L_{\mathrm D},
\end{equation}
where $\mathbf s^{(l)}$ is a pilot symbol of the satellite, $\mathbf w_{\mathrm S}\in\mathbb{C}^{N_{\mathrm S}\times1}$ represents the transmit beamforming vector used at the satellite during training, and $\mathbf n^{(l)}\sim\mathcal{CN}(\mathbf 0,\sigma^2 {\bf I}_{N_{\mathrm G}})$ denotes additive white Gaussian noise.

%Based on the received pilot signals, existing spectral-based algorithms such as MUltiple SIgnal Classification (MUSIC) algorithm can be applied to estimate the AoA pairs~\cite{Schmidt1986MUSIC,Xiong2025Efficient}. Specifically, the array covariance matrix of the received signal can be calculated as
Based on the received pilot signals, the GN forms the sample covariance matrix
\begin{equation}
	 \mathbf{R} = \mathbb{E}\left\{\mathbf y_{\mathrm G}\mathbf y^{\mathrm H}_{\mathrm G} \right\} = \frac{1}{L_{\mathrm D}}\sum_{l=1}^{L_{\mathrm D}}\mathbf y_{\mathrm G}^{(l)}(\mathbf y^{(l)}_{\mathrm G})^{\mathrm H}.
\end{equation}
%After conducting the eigenvalue decomposition, we can obtain
%\begin{equation}
%	\mathbf{R(\Theta)}=\mathbf E_s \mathbf \Sigma_s \mathbf E_s^{\mathrm H}+\mathbf E_n \mathbf \Sigma_n \mathbf E_n^{\mathrm H},
%\end{equation}
%where $\mathbf \Sigma_s \in \mathbb{C}^{1\times1}$ and $\mathbf \Sigma_n \in \mathbb{C}^{(N-1)\times(N-1)}$ are diagonal matrices having the largest and $(N-1)$ smallest eigenvalues of $\mathbf R$ on their diagonal, respectively.
%The matrices $\mathbf E_s \in \mathbb{C}^{N\times1}$ and $\mathbf E_n \in \mathbb{C}^{N\times(N-1)}$ contains the eigenvectors corresponding to eigenvalues on the diagonal of $\mathbf \Sigma_s$ and $\mathbf \Sigma_n$, respectively.
By performing eigenvalue decomposition on $\mathbf R$, the signal and noise subspaces can be obtained. 
Leveraging the orthogonality between these two subspaces, the AoA pair at the GN can be estimated using the MUltiple SIgnal Classification (MUSIC) algorithm~\cite{Schmidt1986MUSIC,Xiong2025Efficient}. Specifically, the pseudo-spectrum function is given by
%By utilizing the orthogonal relationship between the signal subspace and the noise subspace, the spectrum function of signal sources can be presented by
\begin{equation}
	\label{eq:spectrum}
	V(\vartheta,\varphi,\mathbf F_{\mathrm G})=\frac{1}{\mathbf b_{\mathrm G}^{\mathrm H}(\mathbf F_{\mathrm G};\vartheta,\varphi)\mathbf{E}_n\mathbf{E}_n^{\mathrm H}\mathbf b_{\mathrm G}(\mathbf F_{\mathrm G};\vartheta,\varphi)},
\end{equation}
where $\mathbf E_n$ denotes the noise subspace.
The AoA pair $(\hat{\vartheta}_{\mathrm G},\hat{\varphi}_{\mathrm G})$ at the GN side is then obtained by locating the maximum peak of the pseudo-spectrum in \eqref{eq:spectrum}.
%Note that the steering vector $\mathbf a_{\mathrm G}(\vartheta_{\mathrm G},\varphi_{\mathrm G})$ also depends on the pointing matrix $\mathbf{F}_1$ of the RA array, which lies in its array gain pattern term. This implies that an appropriate design of the pointing matrix during the channel-training period is essential for achieving accurate channel estimation.

With the estimated AoA pair, the incident signal direction is reconstructed as
\begin{equation}
    \hat{\mathbf u}_{\mathrm G} = [\sin\hat{\vartheta}_{\mathrm G}\cos \hat{\varphi}_{\mathrm G},\sin\hat{\vartheta}_{\mathrm G}\sin\hat{\varphi}_{\mathrm G},\cos\hat{\vartheta}_{\mathrm G}]^{\mathrm{T}}.
\end{equation}
The GN then updates the boresight directions of all RAs according to the optimal solution in \eqref{eq:optBoresight}.
Subsequently, the beamforming vector $\mathbf w^{\star}_{\mathrm G}$ is obtained by substituting the updated pointing matrix $\mathbf F^{\star}_{\mathrm G}$ and $\hat{\mathbf u}_{\mathrm G}$ into \eqref{eq:arrayRes1} and \eqref{eq:beamform}.

In the uplink training stage, the GN transmits uplink pilots using the optimized beamforming vector $\mathbf w^{\star}_{\mathrm G}$ and pointing matrix $\mathbf F^{\star}_{\mathrm G}$.
%%The GN employs the previously optimized beamforming vector $\mathbf w^{\star}_{\mathrm G}$ and pointing matrix $\mathbf F^{\star}_{\mathrm G}$, while the satellite dynamically adjusts the pointing matrix $\mathbf F_{\mathrm S}$ at the satellite over successive pilot symbols to facilitate the estimation.
%Meanwhile, the satellite varies its RA boresight directions across pilot symbols to facilitate angle estimation.
By applying the same estimation procedure as in the downlink stage, the satellite obtains its estimated AoA pair $(\hat{\vartheta}_{\mathrm S},\hat{\varphi}_{\mathrm S})$, and updates its beamforming vector $\mathbf w^{\star}_{\mathrm S}$ and pointing matrix $\mathbf F^{\star}_{\mathrm S}$ according to \eqref{eq:beamform} and \eqref{eq:optBoresight}, respectively.

\vspace{-0.4cm}
\subsection{Boresight Tracking}
\vspace{-0.1cm}

After the initial angular estimation stage, GN and satellite enter the data transmission stage, during which their local angular parameters may change due to the satellite motion. Re-estimating these parameters through frequent pilot exchange would incur excessive overhead. To address this issue, we exploit the predictability of the satellite trajectory to track the local angular evolution at both nodes with low complexity.

Specifically, let $(\hat{\vartheta}_{\mathrm X}^{[T_0]}, \hat{\phi}_{\mathrm X}^{[T_0]})$ with $\mathrm X \in \{\mathrm{G,S}\}$ denote the local AoA pair estimated at node X at the beginning of a data transmission frame, and let $T_0$ and $T_E$ represent the start and end times of the frame, respectively. Over this short interval, the angular variation is approximated by a linear model. Accordingly, for any $t \in [T_0,T_E]$, the local angular pair at node X is predicted as
\begin{equation}
\hat{\vartheta}_{\mathrm X}^{[t]} = \hat{\vartheta}_{\mathrm X}^{[T_0]} + (t-T_0)\Delta \vartheta_{\mathrm X},
\end{equation}
\begin{equation}
\hat{\varphi}_{\mathrm X}^{[t]} = \hat{\varphi}_{\mathrm X}^{[T_0]} + (t-T_0)\Delta \varphi_{\mathrm X},
\end{equation}
where $\Delta \vartheta_{\mathrm X}$ and $\Delta \varphi_{\mathrm X}$ denote the angular variation rates at node X within the current frame.

Since the satellite orbit is usually known a priori, these angular variation rates can be calculated from the orbital parameters and the relative geometry between the GN and the satellite. 
%Based on the predicted angular pair $(\hat{\vartheta}_{X}^{[t]}, \hat{\varphi}_{X}^{[t]})$, node $X$ updates its propagation direction estimate, and then recomputes the RA boresight directions and beamforming vector according to the closed form solutions in Section III. In this way, both nodes can continuously refine their boresight alignment and beamforming during data transmission without additional pilot signaling.
%
For ease of illustration, we consider the special case where the GN and the satellite orbit lie in the same $y$-$z$ plane. In this case, we have $\varphi_{\mathrm{G}}^{[t]}=\varphi_{\mathrm{S}}^{[t]}=0$, and only the elevation angles need to be tracked. Based on the relative motion geometry, the angular variation rate at the GN can be approximated as
\begin{equation}
\Delta \vartheta_{\mathrm{G}} \approx \frac{\omega L_{\mathrm{O}}}{\bar{d}_{\mathrm{S-G}}},
\end{equation}
where $\omega$ denotes the satellite angular velocity and
\begin{equation}
\bar{d}_{\mathrm{S-G}}=\frac{1}{T_E-T_0}\int_{T_0}^{T_E} d_{\mathrm{S-G}}^{[t]} \, dt
\end{equation}
is the average GN-satellite distance over the considered data frame. 
%The angular variation rates for the other local parameters can be obtained similarly and are omitted for brevity. While other linear increments can also be similarly obtained, those details are omitted for brevity due to space limitation.
The angular variation rates and corresponding linear increments of the remaining local parameters can be derived analogously and are omitted due to space limitations.

\vspace{-0.5cm}
\section{Simulation Results}
\vspace{-0.1cm}
%\begin{figure}[!t]
%	\centering
%	\includegraphics[width=0.8\linewidth]{rateVsTime.eps}
%    \vspace{-0.5cm}
%	\caption{Achievable rate versus time index under perfect CSI.}
%	\label{sim:time}
%    \vspace{-0.5cm}
%\end{figure}

%\begin{figure}[!t]
%	\centering
%	\includegraphics[width=0.8\linewidth]{rateVsDirectivity.eps}
%    \vspace{-0.5cm}
%	\caption{Achievable rate versus antenna directivity under perfect CSI.}
%	\label{sim:p}
%    \vspace{-0.5cm}
%\end{figure}

\begin{figure}[!t]
\centering
	\subfloat[]{
	\includegraphics[width=0.65\linewidth]{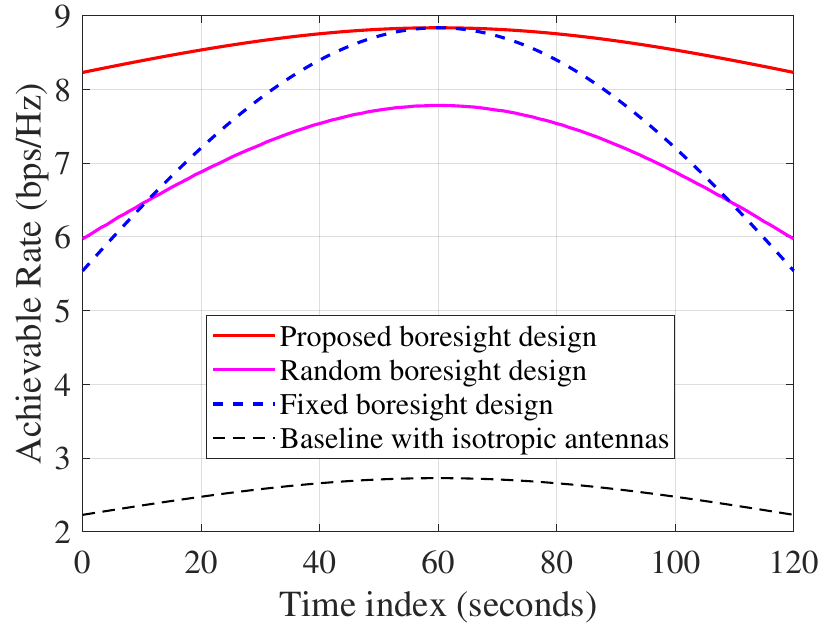}
		\label{sim:time}
	} \\ \vspace{-0.4cm}
	\subfloat[]{
	\includegraphics[width=0.65\linewidth]{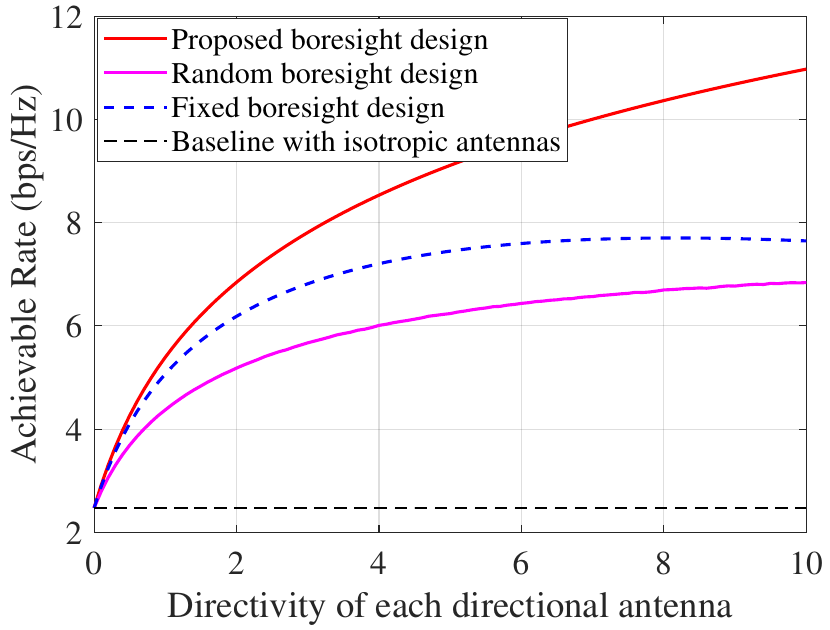}
		\label{sim:p}
	}
	\caption{(a) Achievable rate versus time index under perfect CSI. (b) Achievable rate versus antenna directivity under perfect CSI.}
	\label{fig:achievableRate}
    \vspace{-0.6cm}
\end{figure}
In this section, we present simulation results to validate the performance of our proposed RA-aided LEO satellite communication system with distributed channel estimation and tracking strategy.
In our simulations, we consider a circular LEO orbit with radius $L_{\mathrm O}=6.97\times10^6$ m, corresponding to an orbital period of approximately 96 minutes.
The GN is located at a fixed position of $(0,R_{\mathrm E}+50,0)$ m, where the Earth radius is $R_{\mathrm E}=6.37\times10^6$ m.
Both the GN and the satellite are equipped with $2\times2$ UPA-based RA arrays.
%Therefore, the orbital period is given by $T_{\mathrm S}=5.7879\times10^3 \mathrm{ s} \approx 96$ minutes.
%The initial orbital phase angle at time $t=0$ is set to $\alpha_0 = \alpha(0) = -3.7^{\circ}$.
%Under the considered 3D CCS, we assume that the central (reference) points of the GN is located at the fixed position of $(0,R_{\mathrm E}+50,0)$ m with $R_{\mathrm E}=6.37\times10^6$ m, while the central (reference) point of the satellite at time $t$ is represented by $\left(0, L_{\mathrm O}\cos\vartheta_S(t), L_{\mathrm O}\sin\vartheta_S(t)\right)$ m.
In particular, under the considered setup, we have $\varphi^{[t]}_{\mathrm G}=\varphi^{[t]}_{\mathrm S}=0$, such that the dominant angular variation is captured by $\{\vartheta^{[t]}_{\mathrm G},\vartheta^{[t]}_{\mathrm S}\}$.
%Additionally, the GN and the satellite are equipped with UPAs that consist of
%$N = 2\times2$ and $M = 2\times2$ RAs, respectively. The reference path gain at the distance of 1 m is set as $\beta = -30$ dB for the GN-satellite link, and the noise power at the GN and satellite is set as $\sigma^2_N=-90$ dBm.
%We assume that the satellite communication system operates at the very high frequency (VHF) of 150 megahertz (MHz) with the wavelength of $\lambda=2$ m. Moreover, we set the equal antenna/element spacing at the GN and satellite as $\Delta_{\mathrm G}=\Delta_{\mathrm S}=\lambda/8=0.25$ m.
%
%In addition, we assume that the satellite communication system operates at the very high frequency (VHF) of 150 megahertz (MHz) with the wavelength of $\lambda=0.5$ m.
In addition, the carrier frequency is set to $f_c=2\text{ GHz}$,  corresponding to a wavelength of $\lambda=0.15$ m.
Accordingly, the antenna spacing is $\Delta_{\mathrm G}\hspace{-0.1cm}=\hspace{-0.1cm}\Delta_{\mathrm S}\hspace{-0.1cm}=\hspace{-0.1cm}\lambda/2\hspace{-0.1cm}=\hspace{-0.1cm}0.075$ m, the reference path gain at a distance of 1 m is $\beta = -30$ dB, and the noise power at both the GN and the satellite is $\sigma^2_N=-105$ dBm.

To evaluate the performance of the proposed RA boresight alignment algorithm, we compare it with the following benchmark schemes: 1) Random boresight design, where the boresight direction of each RA is randomly generated within the rotational ranges constrained by \eqref{eq:range}. 2) Fixed boresight design, where the boresight directions of all RAs are fixed at their reference orientations (i.e., the $z$-axis of their local CCS). 3) Baseline with isotropic antennas, where the directional gain is set to $G_{\mathrm{max}}=1$ with $p=0$ in \eqref{eq:pattern}. Note that the above benchmarks differ only in their boresight design. All schemes employ MRT/MRC beamforming and hence rely on the same channel estimation procedure described in Section~\ref{sec:estimation}.

%To evaluate the performance of the proposed RA boresight alignment algorithm, 
Fig.~\subref*{sim:time} 
%Fig.~\ref{sim:time} 
shows the achievable rate at the GN over different schemes for one satellite pass under perfect channel state information (CSI).
%, and compares it with the random boresight design, the fixed boresight design, and the isotropic antenna baseline.
As expected, the proposed boresight design achieves the highest rate among all schemes, as it continuously steers each RA toward the instantaneous satellite/GN direction and thus preserves strong array gain along the satellite trajectory. In contrast, the fixed boresight design performs well only when the satellite moves near its preset pointing direction, and its performance degrades as the angular misalignment increases. 
For the random boresight design, the achievable rate exhibits a similar trend as the fixed boresight scheme but at lower rates.
This confirms that without proper boresight control, the array cannot achieve an adequate directional gain toward the satellite.
The isotropic baseline yields the lowest rate as no directional gain is exploited.
Overall, the results confirm that adaptive boresight control is essential for maintaining high link quality under rapid angular variation.

Fig.~\subref*{sim:p} illustrates the achievable rate at the GN as the antenna directivity $p$ increases to further demonstrate the importance of directional control.
The proposed boresight design consistently achieves the highest rate and continues to benefit from increased directivity by forming a more concentrated beam toward the satellite. 
%Notably, the fixed boresight design also gains from higher directivity at low and moderate values, but its rate saturates once the angular misalignment becomes the dominant limitation.
Both the fixed and random boresight designs benefit from higher directivity at low and moderate values, but their gains become marginal at high directivity because angular misalignment gradually becomes the dominant performance limitation.
The isotropic baseline remains nearly constant since no directional enhancement is introduced. These results show that increasing antenna directivity is beneficial only when supported by proper boresight control.

\begin{figure}[t]
	\centering
	\includegraphics[width=0.7\linewidth]{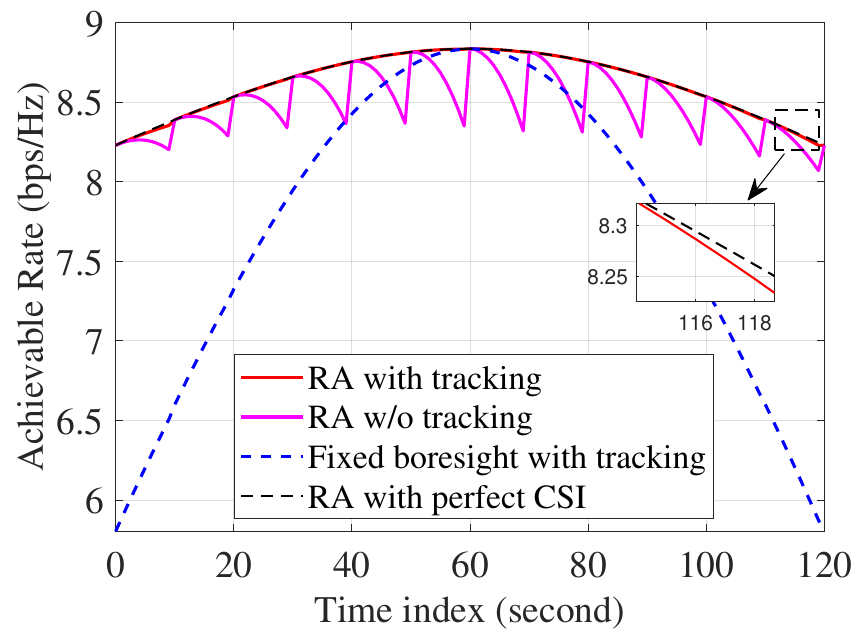}
    \vspace{-0.3cm}
	\caption{Achievable rate versus time index under different schemes.}
	\label{sim:tracking}
    \vspace{-0.7cm}
\end{figure}

To evaluate the proposed channel estimation and tracking algorithm, we consider the previously defined benchmark schemes under different tracking strategies, including: 1) RA without tracking, where the antenna boresight and beamforming are updated by the proposed scheme only during training intervals, and 2) fixed boresight with tracking, where only beamforming is continuously updated while the antenna boresight remains fixed.
The data transmission frame for all schemes has a duration of $T_E-T_0=10$ s.
Fig.~\ref{sim:tracking} illustrates the achievable rate at the satellite over time across different schemes. The proposed RA system with beam tracking nearly matches the perfect CSI upper bound, demonstrating the high accuracy of the proposed angle estimation and tracking procedure. In contrast, the RA system without beam tracking exhibits periodic performance drops as angular misalignment accumulates between updates, highlighting the volatility of high-mobility LEO links. Meanwhile, the fixed boresight baseline with channel tracking provides only moderate performance, as it fails to physically align the high-gain radiation pattern with the satellite. 
%These results confirm that continuous, adaptive beam tracking is essential for sustaining optimal link quality throughout the satellite trajectory.

\vspace{-0.4cm}
\section{Conclusion}
\vspace{-0.1cm}
In this letter, we investigated joint beamforming and boresight optimization for RA-enabled LEO satellite communication. By leveraging the rank-one LoS channel structure, closed-form solutions were derived to enable joint beamforming and boresight direction optimization design with low computational complexity. The proposed design provides an efficient way to exploit rotational DoFs for maintaining directional alignment under high mobility. In addition, an efficient distributed tracking strategy was developed to sustain performance over time. Simulation results demonstrated consistent rate improvements over fixed and random boresight baselines, validating the effectiveness of the proposed framework.

\vspace{-0.3cm}
\bibliographystyle{IEEEtran}
\bibliography{ref}

\end{document}